\documentclass[aip,graphicx,reprint,twocolumn]{revtex4-1}

\usepackage[twocolumn,textwidth=17cm,columnsep=.81cm]{geometry}

\usepackage{graphicx} 
\usepackage{gensymb}
\usepackage{siunitx}
\usepackage[]{lineno}
\newcommand{\iondensity}[2] {#1$\times$10\textsuperscript{#2}\,ions/cm\textsuperscript{2}}

\draft 

\begin{document}


\title{Focused helium ion beam nanofabrication by near-surface
swelling}



\author{Sherry Mo}
\affiliation{Department of Materials Science and Engineering, University of California, Berkeley, CA 94720, USA}


\author{Dana O. Byrne}
\affiliation{Department of Chemistry, University of California, Berkeley, CA 94720, USA}

\author{Frances I. Allen}

\affiliation{Department of Materials Science and Engineering, University of California, Berkeley, CA 94720, USA}
\affiliation{California Institute for Quantitative Biosciences, University of California, Berkeley, CA 94720, USA}


\date{\today}

\begin{abstract}
The focused helium ion beam microscope is a versatile imaging and nanofabrication instrument enabling direct-write lithography with sub-\qty{10}{nm} resolution. Subsurface damage and swelling of substrates due to helium ion implantation is generally unwanted. However, these effects can also be leveraged for specific nanofabrication tasks. To explore this, we investigate focused helium ion beam induced swelling of bulk crystalline silicon and free-standing amorphous silicon nitride membranes using various irradiation strategies. We show that the creation of near-surface voids due to helium ion implantation can be used to induce surface nanostructure and create subsurface nanochannels. By tailoring the ion dose and beam energy, the size and depth of the swollen features can be controlled. Swelling heights of several hundred nanometers are demonstrated and for the embedded nanochannels, void internal diameters down to \qty{30}{nm} are shown. Potential applications include the engineering of texturized substrates and the prototyping of on-chip nanofluidic transport devices. \vspace{15mm}

\end{abstract}


\maketitle 



\section{Introduction}

Focused ion beam (FIB) microscopes are routinely used for micro- to nanoscale material removal by sputtering, e.g.\ to create cross-sections for 3D volume imaging by scanning electron microscopy (SEM)~\cite{Berger2023} and for the preparation of specialized sample geometries for transmission electron microscopy (TEM)~\cite{Mayer2007} and atom probe tomography (APT)~\cite{Prosa2017}. Beyond these sample preparation tasks, FIB instruments are increasingly being used for many other applications, including analytics by secondary ion mass spectrometry (SIMS) and device prototyping. These advancements have been propelled by the wide range of ion species that today's FIB sources can deliver~\cite{Hoflich2023}. Notably, a number of emerging FIB applications leverage particular ion irradiation effects that are unwanted and minimized for the more conventional applications like sample preparation. For example, the creation of a near-surface damage layer and the implantation of ions into the specimen over the ion stopping range is not desired in TEM lamella preparation. However, in other applications, these beam damage effects are specifically targeted, such as to tune material properties through local amorphization~\cite{Allen2021} and to selectively dope materials with specific species such as silicon atoms to form single photon emitters~\cite{Schroder2017,Hollenbach2022}.

Ion implantation from gaseous species such as hydrogen and helium can lead to the formation of nanobubbles and larger gas-filled voids in the target material~\cite{Primak1966}. The mechanism of nanobubble formation is understood to involve diffusion and recombination processes of the vacancy defects and interstitial gas atoms that are produced during the ion collision cascade. As the ion dose increases, nanobubbles coalesce to form larger voids and eventually a sub-surface crack plane can develop, leading to exfoliation of the top layer of the material~\cite{Allen2020}. On the one hand, fundamental understanding of these irradiation effects is important for the design of radiation-tolerant materials. In fact, recently helium FIBs have been used to conduct site-specific and dose-controlled systematic studies of nanobubble formation and blistering in a range of materials~\cite{Allen2020,Livengood2009,Wang2016,Veligura2013,Chen2020}
However, these swelling based ion implantation effects can also be leveraged for nanofabrication tasks, which is the focus of the present study.

The nanoscale positioning accuracy and dose control of the helium FIB has enabled the exploration of helium ion induced swelling for advanced 3D nanofabrication, as demonstrated by several groups. For example, silicon substrates have been irradiated with helium ions to form raised lines with a half-pitch of \SI{3.5}{nm}, and in the same work, concentric square and circular irradiations were used to create raised 3D pyramid and cone-shaped structures~\cite{Zhang2015}. 
In another study, free-standing silicon and diamond lamellae were irradiated with helium ions to form longitudinal and hemispherical nanoscale surface protrusions, depending on the scanning strategy used (line vs.\ spot mode)~\cite{Kim2019}.
Focused helium ion irradiation and swelling of silicon substrates over broader areas has also been demonstrated as a method to deform supported structures, such as nanoporous aluminum oxide films, leading to the enlargement of the nanopores in the film due to the substrate-induced strain~\cite{Aramesh2018}.

In that which follows, we continue explorations of nanofabrication by helium ion induced swelling, using the helium FIB to create texturized surfaces and buried nanochannels in both bulk silicon substrates and free-standing silicon nitride membranes. We show how the size of the near-surface voids that are responsible for the local surface deformation vary predictably with ion dose and how void/nanochannel depth can be controlled via the ion beam energy. Examples of various patterning geometries illustrate the versatility and tunability of this single-step nanofabrication technique, with potential applications that include surface texturing for strain engineering and nanoimprint lithography, and the fabrication of embedded nanochannels for nanofluidic device prototyping.

\section{Experiment}

\subsection{Sample Preparation}
For the tests on bulk substrates, silicon (111) chips were cleaned by rinsing with acetone followed by isopropyl alcohol before drying using compressed air. The free-standing membrane samples were \qty{200}{nm} thick silicon nitride (SiN) (Ted Pella) on \qty{3}{mm} diameter silicon frames, that were sputter-coated on the back (frame) side with gold-palladium to an approximate thickness of \qty{3.5}{nm} and mounted topside down for helium ion irradiation. A similar silicon nitride chip with pre-fabricated \qty{2}{\micro\meter} holes was prepared in the same manner. Topside down mounting ensured that the free-standing portion of the membrane could easily be discerned from the supporting frame when imaged from above.

\subsection{Irradiation and Imaging}
All samples were irradiated using a Zeiss ORION NanoFab helium ion microscope (HIM) under normal incidence at beam energies ranging from \SIrange{10}{20}{keV}, depending on the ion penetration depth desired. Beam currents were typically around \SIrange{3}{4}{pA} using a \SI{20}{\micro\meter} aperture, spot control 4, and helium gas pressure at the source of 2$\times$10$^{-6}$\,Torr. Irradiation doses mainly focused on the ranged from \iondensity{6}{17} to \iondensity{12}{17}. Depending on the pattern size and target dose, individual features took between 1 and 30\, minutes to irradiate.

Various irradiation patterns were implemented using Nanopatterning and Visualization Engine (NPVE) software from Fibics, Inc. Circular and annular pattern arrays were defined using the in-built NPVE patterning functions with raster scan and concentric scan routines.  
For the bulk silicon samples, the circle patterns had a diameter of \SI{1}{\micro\meter} and the annuli had inner and outer diameters of \SI{0.6}{\micro\meter} and \SI{1}{\micro\meter}, respectively. 
An additional annulus with an inner and outer diameter of \SI{1.2}{\micro\meter} and \SI{2}{\micro\meter}, respectively, received a dose of \iondensity{8}{17}; the larger dimensions of this annulus were chosen for preliminary gallium FIB cross section analysis of the embedded nanochannel. The annuli patterned into the free-standing membrane samples had inner and outer diameters of \SI{2.4}{\micro\meter} and \SI{3.4}{\micro\meter}, respectively. These larger annulus dimensions were chosen so that the swelling features would circumscribe the prefabricated \qty{2}{\micro\meter} holes in the particular silicon nitride substrate used. Separately, a set of \qty{0.5}{\micro\meter} by \qty{2}{\micro\meter} rectangles were irradiated with \iondensity{8}{17} at three different beam energies: \qty{10}{keV}, \qty{15}{keV}, and \qty{20}{keV}. This pattern was implemented on the continuous free-standing silicon nitride substrate to probe the effect of beam energy on ion implantation depth. 

More complex patterns included a Penrose tile and a mock-up of a nanofluidic DNA sorting device. The Penrose tile was imported into NPVE as a bitmap and the sorting device was made with merged shapes in NPVE. All shapes were patterned using a beam dwell time of \SI{1}{\micro\second} and a scan spacing of \qty{0.5}{nm} (nominal probe size \qty{\sim1}{nm}).

HIM imaging of the 3D features that formed as a result of the helium ion irradiation were acquired in-situ at \qty{25}{keV} using a secondary electron detector. Imaging doses were typically \iondensity{4.4}{10} per frame. The HIM images were acquired both under normal incidence and using a stage tilt of \qtyrange{45}{54}{\degree} in order to visualize the raised nature of the features. In addition, the gallium FIB column installed on the ORION NanoFab instrument was used to obtain cross-section views through the patterned features. The gallium ion beam energy used for milling the cross sections was \qty{30}{keV} and the gallium ion beam current varied between \qtyrange{5}{15}{pA}. The mills were monitored periodically by HIM imaging and stopped once the voids inside the swelling features were revealed. 

To further characterize swelling heights, atomic force microscopy (AFM) was performed using a CoreAFM from NanoSurf. The tip used was an AppNano SPM probe (model ACLA), with a cantilever length of \SI{225}{\micro\meter} and width of \SI{40}{\micro\meter}. The operating frequency for contact mode ranged from \SIrange{160}{225}{\kilo\hertz}, with most measurements conducted at around \SI{167}{\kilo\hertz}. Scan sizes were typically around \SI{20}{\micro\meter}, implementing 256 lines at \qty{0.7}{s} per line. The AFM data were post-processed using open-source Gwyddion software, extracting line profiles of the measured heights for various dose series.

\begin{figure*}
    \centering
    \label{fig: Fig. 1}
    \includegraphics[width=1\linewidth]{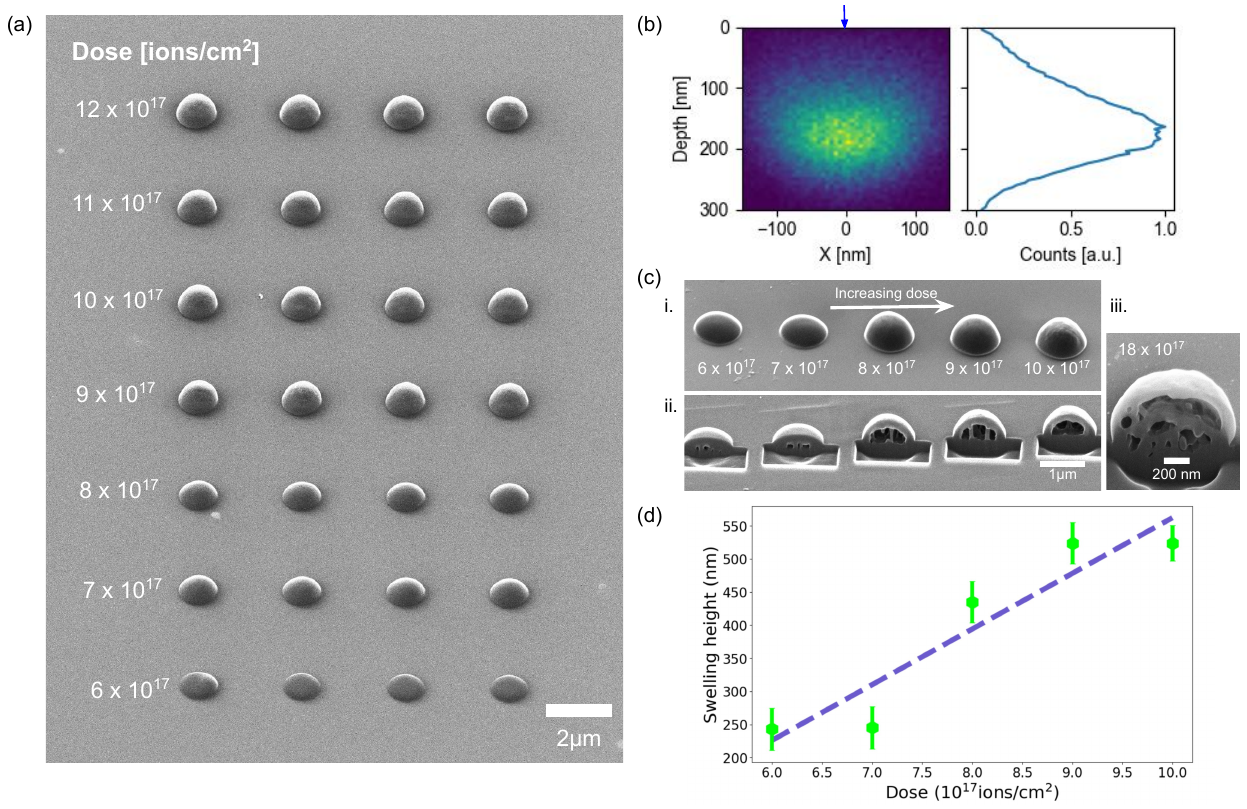}
    \caption{Texturizing surfaces by helium ion induced swelling: (a) Tilted view (\qty{54}{\degree}) of a dose series of \qty{1}{\micro\meter} diameter circular irradiations on bulk silicon for \qty{15}{keV} helium ions. The four features in each row were irradiated to the same dose demonstrating the consistency of swelling height. (b) SRIM simulation of ion stopping range for \qty{15}{keV} helium ions incident on silicon. Left: 2D histogram of final ion positions versus depth and lateral scatter; arrow indicates direction and position of ion incidence. Right: corresponding 1D histogram showing total counts of stopped helium ions versus depth. (c)(i) A series of circular irradiations with doses ranging from 6$\times10^{17}$ to 10$\times10^{17}$\,ions/cm\textsuperscript{2} using \qty{15}{keV} helium ions (ii) Gallium FIB milled cross section views of the same features. (iii) Gallium FIB milled cross section view of a \qty{1}{\micro\meter} circular feature that received \iondensity{18}{17}. (d) A scatter plot of swelling height for the five features in (c) vs.\ dose as measured by AFM. Each data point is the average value obtained for four features of the same dose. Swelling height appears to vary linearly with irradiation dose over the given dose range.}
\end{figure*}

\section{Results and Discussion}

Fig.\ 1a shows a tilted HIM view of a dose series from \iondensity{6}{17} to \iondensity{12}{17} for \qty{15}{keV} helium ions irradiated in circular patterns onto the bulk silicon (111) substrate. In each row the dose was kept fixed, demonstrating the consistency in shape and swelling height of the patterned features in each case. Swelling height increases with dose, and for the circular patterns chosen here, smooth dome-shaped protrusions were formed. 

Simulations of the stopping range of \qty{15}{keV} helium ions in silicon using Stopping Range of Ions in Matter (SRIM) code~\cite{Ziegler2010} indicate that at this beam energy the ions implant shallowly, mostly accumulating at a depth of \qtyrange{100}{200}{\nano\meter} (Fig.\ 1b). While the HIM instrument is more typically operated at higher beam energies (\qtyrange{25}{30}{keV}), it was found that the lower beam energy and associated shorter penetration depth was more effective for local deformation of the silicon surface by swelling. 

Visualization of the inside of the swollen features is shown in Fig.\ 1c, where we present a dose series from \iondensity{6}{17} to \iondensity{10}{17} (Fig.\ 1c(i)) together with cross-section views through the same features created by gallium FIB milling (Fig.\ 1c(ii)). Again, an increase in swelling height with dose is revealed. The cross-section views show that the swollen features are hollow. Furthermore, the silicon shell structures capping the voids appear robust, remaining structurally sound after milling with the gallium ion beam. However, at doses above \iondensity{$\sim$1.4}{18}, the shell around the circular protrusions starts to become irregular. Cross sectioning reveals that the void starts to backfill with additional silicon swelling up from below. This is shown in the cross-section view in Fig.\ 1c(iii) for an irradiation dose of \iondensity{1.8}{18}.  
\begin{figure*}
\centering
\label{fig: Fig. 2}
\includegraphics[width=1\linewidth]{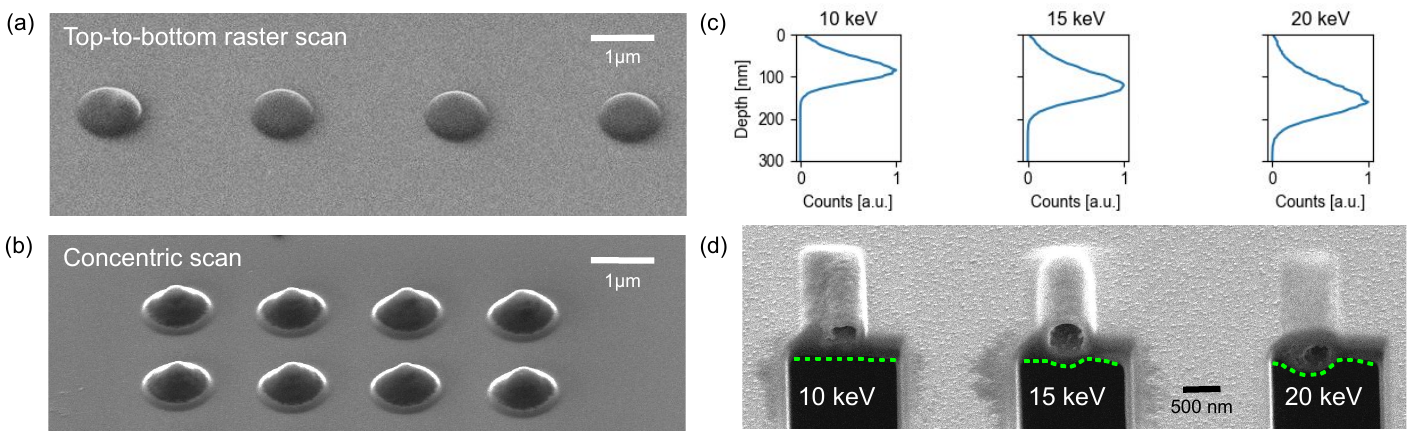}
\caption{Effect of scan strategy and ion beam energy on the surface morphology and ion implantation depth, respectively: (a) Tilted view (\qty{54}{\degree}) of \SI{1}{\micro\meter} diameter circular irradiations using \qty{15}{keV} helium ions at a dose of 6$\times 10^{17}$\,ions/cm$^{2}$ following a top-to-bottom raster scan routine. (b) Tilted view under the same angle for the same dose and beam energy as in (a) for a concentric inside-to-outside scanning routine. (c) SRIM plots of ion stopping range for \qty{10}{keV}, \qty{15}{keV}, and \qty{20}{keV} helium ions incident on silicon nitride. (d) Tilted  view (\qty{45}{\degree}) of a freestanding \qty{200}{nm} silicon nitride membrane that had been irradiated with 8$\times$10\textsuperscript{17}\,ions/cm\textsuperscript{2} at \qty{10}{keV}, \qty{15}{keV}, and \qty{20}{keV} to fabricate linear subsurface channels (using rectangular patterns, width \qty{0.5}{\micro\meter}, length \qty{2}{\micro\meter}). Gallium FIB milling reveals the cross section views. Outlines of the bottom edge of the cross sections show how the channel depth increases with beam energy.}
\end{figure*}
AFM height profile results for circular irradiations of the same diameter using \qty{15}{keV} helium ions for doses from \iondensity{6}{17} to \iondensity{10}{17} are presented in Fig.\ 1d, suggesting a positive linear relation between ion dose and swelling height over this dose range. At higher dose, the increase in swelling height starts to plateau and the swollen surface becomes irregular, as described above. 
Therefore, in order to achieve smooth and predictable surface texturing, helium ion doses in the linear swelling height range were chosen. 

The mechanism for void formation upon helium ion irradiation involves the formation and coalescence of helium nanobubbles~\cite{Livengood2009,Allen2020}. The nanobubbles are believed to be produced when the implanted interstitial helium atoms combine with the ion-induced vacancy defects in the crystal lattice, whereby the highest concentration of nanobubbles forms at around the peak in the ion stopping range. Helium ion irradiation at an energy of \qty{15}{keV}, as in Fig.~1, results in the accumulation of implanted species relatively close to sample surface, which facilitates texture modification on the beam-facing side. As shown in previous HIM studies, the critical dose for helium nanobubble formation in silicon is around \iondensity{1}{17}, with larger voids forming at increasing dose~\cite{Livengood2009}, commensurate with the dependence of swelling height on dose determined in the present study. 

In addition to controlling the height of the surface protrusions via the irradiation dose, the overall shape of the raised features can be controlled by varying the scan strategy. An example of this is shown in Figs.\ 2a and 2b, where the circular irradiations were performed using a line-by-line top-down raster scan routine (same as in Fig.\ 1) vs.\ a concentric inside-to-outside scanning routine. The top-down raster scan resulted in dome-shaped features, whereas the concentric scan starting in the center resulted in cone-shaped features. We speculate that the cause of this variation in swelling morphology is radial diffusion and a preference for helium segregation to grow bubbles where dose accumulates first. With the center-out concentric scanning, the beam dwells in the center first, and therefore causes the cone-like shape. Further results from concentric scanning for various doses are shown in the Supplementary Fig.~S1. 

Another parameter to tune to customize the swelling-induced substrate deformation is the ion beam energy. As mentioned previously, we found that for creating surface texture on the bulk silicon substrates it was beneficial to choose a lower beam energy (\qty{15}{keV}) than the HIM instrument is usually operated at, since this produces the swelling void closer to the surface. In the case of thin membrane substrates, it can be especially useful to delicately tune the beam energy to control the depth at which the void forms, e.g.\ to symmetrically embed the void. To illustrate, Fig.\ 2c shows SRIM simulation results of the ion stopping range for \qty{10}{keV}, \qty{15}{keV} and \qty{20}{keV} helium ions incident on a silicon nitride target. The increase in ion penetration depth with increasing beam energy is apparent. In an experiment, these three beam energies were then used to irradiate rectangular bar shapes into a \qty{200}{nm} thick free-standing silicon nitride membrane (Fig.\ 2d). Each bar received the same dose (\iondensity{8}{17}) and resulted in the formation of a linear channel embedded inside the membrane. Gallium FIB milling was used to obtain cross-section views through the channels to gauge the depth of each.

\begin{figure*}
\centering
\label{fig: Fig. 3}
\includegraphics[width=1\linewidth]{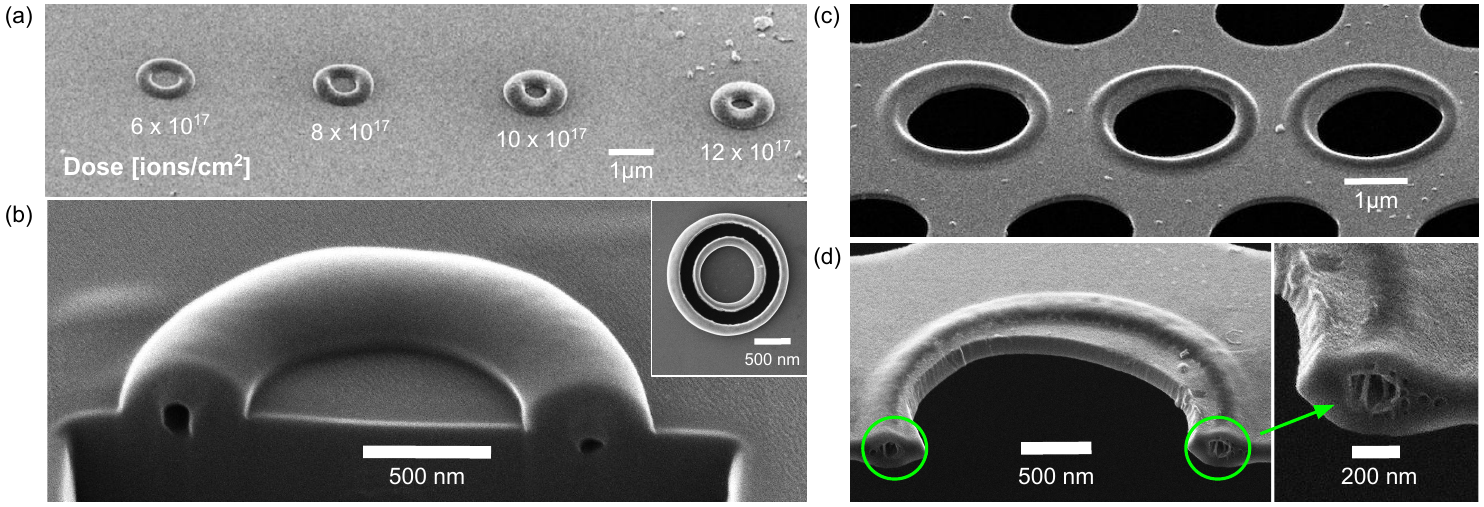}
\caption{Fabricating subsurface nanochannels in bulk and free-standing membrane substrates by helium ion induced swelling: (a) Tilted view (54\degree) of a series of annulus patterns (inner diameter \qty{0.6}{\micro\meter}, outer diameter \qty{1}{\micro\meter}) in bulk silicon irradiated using \qty{15}{keV} helium ions to doses of 6$\times$10\textsuperscript{17} to 12$\times$10\textsuperscript{17}\,ions/cm\textsuperscript{2}.
(b) Tilted  view (same angle) showing a gallium FIB milled cross-section view through a larger annulus (inner diameter \qty{1.2}{\micro\meter}, outer diameter \qty{2}{\micro\meter}) fabricated using a helium ion dose of \iondensity{8}{17}. The inset shows a zoomed in top-down image of a separate annulus that received a dose of \iondensity{7}{17} with the same dimensions as in (a) that was milled open by gallium FIB revealing a continuous void inside. (c) Tilted view (54\degree) of a set of annuli in a free-standing holey silicon nitride membrane (inner diameter \qty{2.4}{\micro\meter}, outer diameter \qty{3.4}{\micro\meter}) fabricated using \qty{15}{keV} helium ions to the same dose of \iondensity{12}{17}. (d) Gallium FIB milled cross-section view through one of the annuli revealing the nanochannel formed by helium ion induced swelling in the silicon nitride membrane. All annulus patterns were scanned concentrically from center outwards.}
\end{figure*}

In the case of the \qty{10}{keV} helium ion irradiation of the silicon nitride membrane shown in Fig.\ 2d, the nanochannel formed closer to the top surface due to the shallower ion implantation depth. The cross-section view indicates that for this lower beam energy the shell around the top of the channel is not evenly dome shaped (i.e.\ the top seems flat), suggesting lower structural integrity. This could be another direct result of the shallower implantation, which could have enabled some of the helium to diffuse to the surface creating the unevenly textured top surface seen in the image. The \qty{15}{keV} irradiation resulted in a channel forming nearer to the center of the membrane, with more symmetric swelling of the top and bottom membrane surface (the dashed lines in the figure outline the deformation of the lower surface to clarify this point). The \qty{15}{keV} channel also has a much more regular circular cross-section, indicating even swelling and channel walls that are sufficiently robust. For the irradiation at \qty{20}{keV}, the channel formed closer to the bottom surface of the membrane. Despite being irradiated at the same dose, this void was smaller in diameter than the \qty{15}{keV} channel, which may be due to a fraction of the ions passing straight through the membrane and thus not implanting at this beam energy, as predicted by the SRIM simulation.

We note that the silicon nitride membrane material used here was non-stoichiometric and therefore amorphous by nature. To date, most HIM swelling studies have focused on crystalline targets, with amorphous materials such as amorphous silicon dioxide being selected as substrates to, in fact, mitigate swelling effects~\cite{Allen2021}. Future work investigating the variable swelling behavior of amorphous target materials would thus be of great interest.

To further explore the fabrication of subsurface nanochannels using the helium ion swelling technique, we performed a series of annulus irradiations of bulk silicon and a free-standing silicon nitride membrane. A helium ion beam energy of \qty{15}{keV} was used throughout. Fig.\ 3a shows a dose series for annulus patterns irradiated onto bulk silicon from \iondensity{6}{17} to \iondensity{12}{17}. The resulting raised annulus features hold true to the positive correlation between dose and swelling height observed in Fig.\ 1. In Fig.\ 3b, a gallium FIB cross-section view through an annulus irradiated into silicon at a dose of \iondensity{8}{17} is shown, revealing an embedded channel of diameter \qty{\sim100}{nm}. In the inset, a further cross-section view is shown, where the gallium FIB was used to mill away the top surface of the annulus, revealing that the channel structure within formed a continuous void throughout the entire structure. The channel walls are expected to be amorphous due to ion damage from the collision cascade~\cite{Livengood2009,Kim2019}.

\begin{figure*}
\centering
\label{fig: Fig. 4}
\includegraphics[width=1\linewidth]{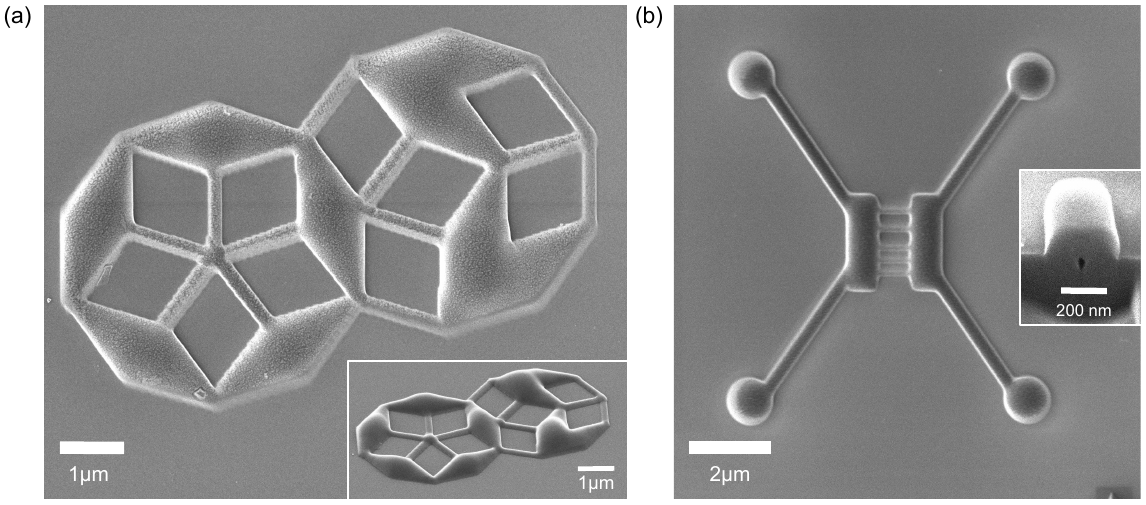}
\caption{More complex surface texture and nanochannel patterning examples on bulk silicon using \qty{15}{keV} helium ions: (a) A Penrose tile that received \iondensity{6}{17}.
The inset shows a tilted view.
(b) A mockup of a sorting device for protein-bound DNA with subsurface nanochannels patterned using a uniform dose of \iondensity{6}{17}. The inset shows the smallest subsurface channel achieved in the present study. The channel width measured here was \qty{30}{nm} for a dose of \iondensity{1.0}{18} (further examples for different doses and pattern widths are shown in Supplementary Fig.~S2).}
\end{figure*}

In Fig.\ 3c, we show the results of annulus irradiations of the free-standing silicon nitride membrane. These annuli were patterned to encircle a set of prefabricated holes in the membrane. A gallium FIB cross-section through one of the annuli is shown in Fig.\ 3d, again revealing the channel structure embedded in the membrane. In the magnified view, one can see how the choice of \qty{15}{keV} for the beam energy resulted in the formation of the channel centered at a depth of \qty{\sim100}{nm} in the \qty{200}{nm} thick membrane. The diameter of the embedded channel is \qty{\sim200}{nm}, resulting in obvious surface protrusions on top and bottom surfaces.

More complex patterning examples are shown in Fig.\ 4. The substrate was again bulk silicon and the helium ion beam energy was \qty{15}{keV}. In the first example (Fig.\ 4a) we show surface texturing in the form of a Penrose tile, created using a uniform dose of \iondensity{6}{17}. 
The tilted view in the inset shows the resulting variation in surface swelling height, revealing more swelling in the case of the wider patterned features. There is also an extra bump at the intersection of the five thin connecting lines in the left portion of the design. Neither effect results from a variation in local irradiation dose, since this was kept constant across the pattern. 
In the Supplementary Information (Fig.~S2), we show additional results from an experiment where lines of increasing width were irradiated at the same dose, confirming the dependence of swelling height on feature width seen in the Penrose tile -- i.e.\ that the widest features swell the most. This effect can be attributed to the relative ease of a larger diameter shell buckling as a result of the helium gas pressure buildup in the void. 
In the case of the increased swelling observed at the node of the five connecting narrow lines, this could be be due to radial diffusion into the center and can be used to fabricate more voluminous nanofluidic nodes. Alternatively, if a continuous swelling height across the feature is desired, one can reduce the dose at the node region.

Figure 4b presents a layout for a nanofluidic sorting device for protein-bound DNA~\cite{Wang2005}, fabricated using a uniform irradiation dose of \iondensity{6}{17}. In the inset, a cross-section view of a test nanochannel segment is also shown, revealing a channel width of \qty{30}{nm}. For the latter, an irradiation dose of \iondensity{1.0}{18} was used. We note that other fabrication approaches for embedded nanochannels using the HIM have also been reported, using resist-based lithography~\cite{Cai2018} and local amorphization of silicon followed by wet-etch and metal layer deposition steps~\cite{Wen2022}. However, the helium FIB swelling technique demonstrated here offers the benefit of being a single-step process, which may facilitate the prototyping of advanced nanofluidic devices. 
For example, one might use the swelling approach to integrate larger on-chip microfabricated channels with nanochannels, offering design possibilities beyond the reach of current conventional methods.

\section{Conclusion}

Highly customizable surface nanostructuring and subsurface nanochannel formation has been demonstrated using the helium FIB swelling method. The swelling phenomenon results from shallow implantation of the gaseous ion species into the target, forming near-surface voids. As such, the helium ion beam energy can be used to tune the depth at which the voids form, e.g.\ \qty{15}{keV} helium ions can be used to form a void centered at a depth of \qty{\sim100}{nm} in a \qty{200}{nm} thick silicon nitride membrane. 
Helium ion doses that ensured smooth surface swelling were identified, e.g.\ for \SI{1}{\micro\meter} wide protrusions patterned on silicon at \qty{15}{keV},  \iondensity{6}{17} to \iondensity{10}{17} gave a consistent increase in swelling height with dose. Swelling heights of several hundred nanometers were achieved. 

To investigate the creation of voids to form nanochannels, line and annulus patterns were chosen. Cross-sectioning of the resulting swollen structures by gallium FIB milling revealed the presence of continuous embedded nanochannels. The precise nanopatterning capability of the helium FIB also enabled the fabrication of more complex designs, such as a Penrose tile (for creating surface texture) and a replicated design of a nanofluidic sorting device (targeting embedded nanochannels). 

As a single-step method applicable to various substrates, the helium FIB swelling technique can be used to efficiently fabricate a range surface topographies and embedded void/nanochannel designs to prototype e.g.\ molds for nanoimprint lithography, substrate surface topographies for strain engineering, and on-chip nanofluidic devices. Further work is needed to determine the minimum achievable nanochannel diameters, investigate the surface roughness of the nanochannel walls, and explore the maximum achievable nanochannel depths. The maximum achievable size of the patterned areas and length (aspect ratio) of the nanochannels will ultimately be determined by the beam current, beam stability, and the stability of the stage.

\begin{acknowledgments}
This work was funded in part by NSF Award No.\ 2110924. D.O.B. also acknowledges funding from the Department of Defense through the National Defense Science \& Engineering Graduate (NDSEG) Fellowship Program. Ion irradiation was performed at the qb3-Berkeley Biomolecular Nanotechnology Center. 
\end{acknowledgments}





%

\renewcommand\thefigure{S\arabic{figure}}    
\setcounter{figure}{0}
 \begin{figure*}[h]
 \centering
\includegraphics[width=0.9\textwidth]{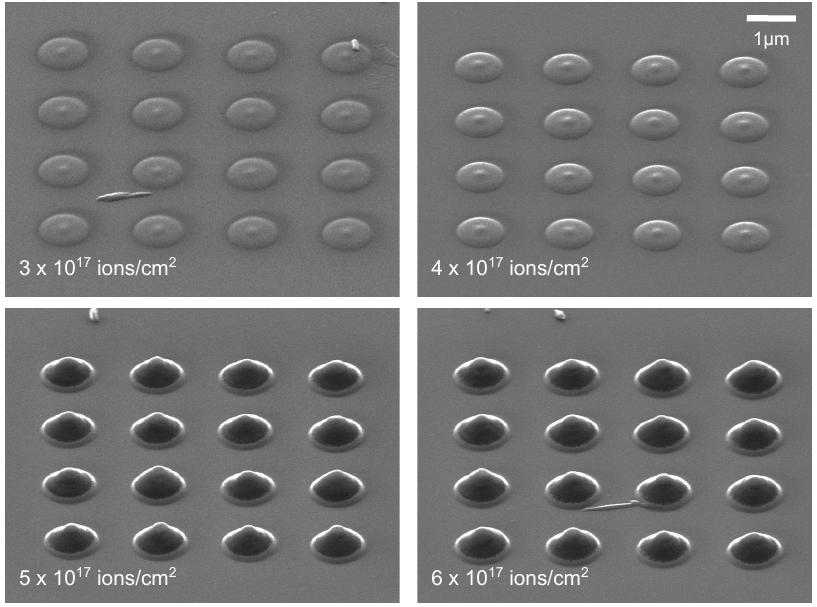}
 \caption{Tilted views (\qty{54}{\degree}) of \SI{1}{\micro\meter} diameter circular irradiations using \qty{15}{keV} helium ions following a concentric inside-to-outside scanning routine. Four irradiation doses were tested, as labeled in the images.}
\label{Fig:Concentric_scan_dose_series}
\end{figure*}

\begin{figure*}[h]
 \centering
\includegraphics[width=0.9\textwidth]{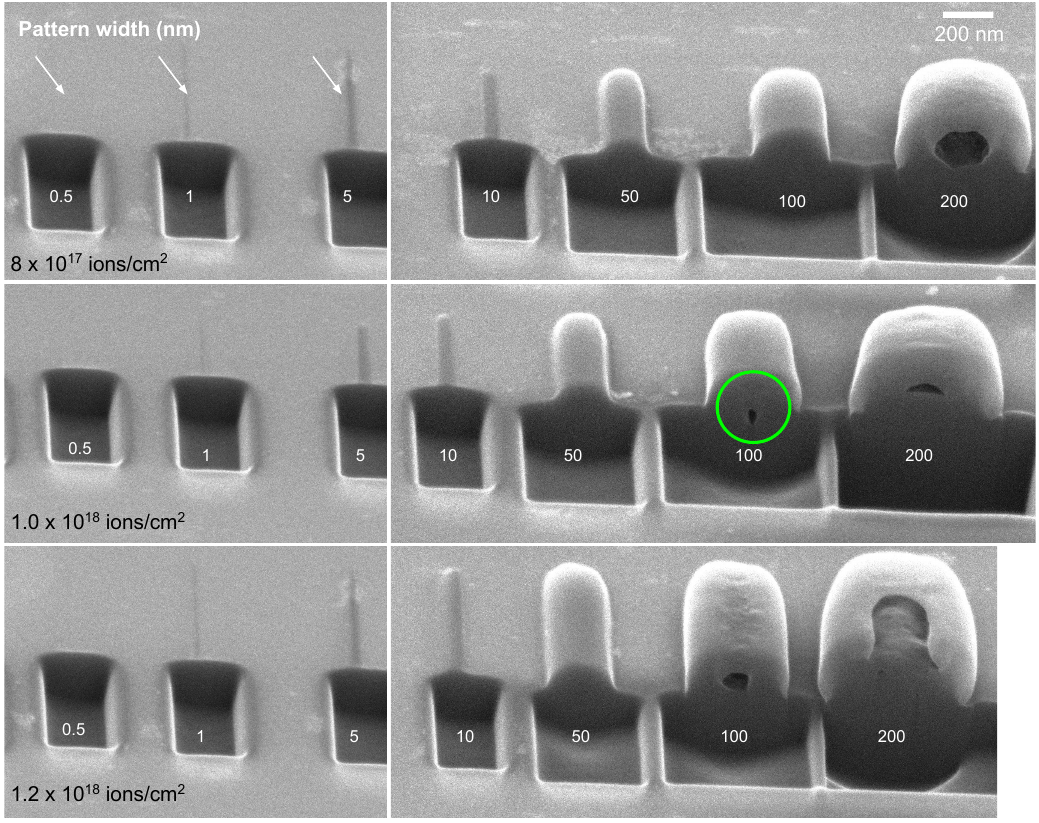}
 \caption{Tilted views (\qty{54}{\degree}) of gallium FIB milled cross-sections through helium ion irradiations for a range of nominal pattern widths and doses. Swelling height increases with increasing pattern width for a given dose. The highlighted nanochannel (green circle) is the smallest produced in the current study and has a diameter of \qty{\sim30}{nm}.}
\label{Fig:Width_test}
\end{figure*}

\end{document}